\documentclass[aip,jmp, amsmath,amssymb,reprint,]{revtex4-1}

\usepackage{graphicx}
\usepackage{dcolumn}
\usepackage{bm}
\usepackage{color}

\begin{document}
\title{A hybrid polarization-selective atomic sensor for radio-frequency field detection with a passive resonant-cavity field amplifier}

\author{D.~A.~Anderson}
\email{dave@rydbergtechnologies.com.}
\affiliation{Rydberg Technologies, Ann Arbor, Michigan 48104 USA}
\author{E.~G.~Paradis}
\affiliation{Rydberg Technologies, Ann Arbor, Michigan 48104 USA}
\affiliation{Eastern Michigan University, Ypsilanti, Michigan, 48197 USA}
\author{G.~Raithel}
\affiliation{Rydberg Technologies, Ann Arbor, Michigan 48104 USA}
\affiliation{Department of Physics, University of Michigan, Ann Arbor, Michigan 48109 USA}

\date{\today}

\begin{abstract}
We present a hybrid atomic sensor that realizes radio-frequency electric field detection with intrinsic field amplification and polarization selectivity for robust high-sensitivity field measurement.  The hybrid sensor incorporates a passive resonator element integrated with an atomic vapor cell that provides amplification and polarization selectivity for detection of incident radio-frequency fields. The amplified intra-cavity radio-frequency field is measured by atoms using a quantum-optical readout of AC level shifts of field-sensitive atomic Rydberg states.  In our experimental demonstration, we employ a split field-enhancement resonator embedded in a rubidium vapor cell to amplify and detect C-band radio-frequency fields. We observe a field amplification equivalent to a 24~dB gain in intensity sensitivity. The spatial profile of the resonant field mode inside the field-enhancement cavity is characterized. The resonant field modes only couple with a well-defined polarization component of the incident field, allowing us to measure the polarization of the incident field in a robust fashion. Measured field enhancement factors, polarization-selectivity performance, and field distributions for the hybrid sensor are in good agreement with simulations.  Applications of hybrid atomic sensors in ultra-weak radio-frequency detection and advanced measurement capabilities are discussed.
\end{abstract}

\maketitle
\section{Introduction}

The emergence of quantum sensor technologies is driving a paradigm shift in modern sensing and measurement instrumentation by enabling fundamentally new detection capabilities and performance metrics that are unmatched by those of traditional sensor technology~\cite{Ludlow.2015,Taylor.2008,Patton.2012,Battelier.2016}.  Quantum sensing of radio-frequency (RF) electric fields~\cite{Sedlacek.2012} using Rydberg electromagnetically-induced transparency (EIT) in atomic vapors~\cite{Mohapatra.2007} has made rapid progress towards viable quantum RF measurement technologies~\cite{Sedlacek.2013,Holloway.2014,Anderson.2016}.  Notable advances include the realization of compact sensing elements capable of broadband RF measurement from MHz~\cite{Miller.2016} to millimeter-wave~\cite{Gordon.2014} and THz electric fields~\cite{Wade.2016}, detection of weak fields at the $\sim$0.1~mV/m level~\cite{Kumar.2017}, to measurements of high-intensity $>$1~kV/m fields with continuous RF frequency tuning~\cite{Anderson.2017}.  Demonstrated applications of Rydberg sensors have included sub-wavelength near-field and far-field imaging of antenna radiation patterns, mapping of microwave circuitry, characterizations of large RF systems such as gigahertz transverse electromagnetic (GTEM) cells~\cite{HollowayEMC.2017,Anderson.2018}, and measurements of RF noise~\cite{Hollowaynoise.2017}.

Several limitations still need to be overcome for the Rydberg-EIT RF field measurement method to be applied in practical quantum RF field sensors.  These include improving the achievable sensitivity: the most sensitive measurements demonstrated to date have measured fields down to the $\sim$0.1~mV/m level~\cite{Kumar.2017}, limited primarily by EIT line width and shot noise of the optical readout.  Further, this sensitivity level has been achieved by detecting small changes in the EIT peak line shape.  Extraction of the RF electric field from a detailed analysis of the EIT line shape requires a complex model that depends on experimental parameters such as excitation beam powers, diameters, and vapor pressures.  These dependencies may affect the reliability of operation and the suitability as an absolute field measurement method.

In the present work we explore passive electric-field amplification as a means to achieve sensitive field detection with the robustness that is inherent to RF field measurements based on EIT spectroscopy of field-induced Rydberg energy-level splittings or shifts. We realize this goal by using a hybrid atomic sensor that combines the advantages of Rydberg-based quantum RF field measurement with those of integrated RF resonator structures.  The demonstrated resonators are solid metal structures that enable passive, polarization-selective compression and amplification of incident RF fields in an electric-field mode volume as small as a few cubic millimeters, for resonant RF wavelengths of several centimeters. While the resonators mimic the function of traditional radio receiver LC circuits, their small size allows for integration inside cm- to mm-sized spectroscopic vapor cells. As the field amplification is based on passive electric-field buildup in a rigid metal resonator, the hybrid atomic sensor enables robust atom-based RF field sensing by measuring spectroscopic level shifts and splittings that are comparatively insensitive to experimental parameters.

%afford exquisite control and conditioning of the RF field measured by the atomic states to achieve passive RF field amplification and polariziation sensitivity for robust atom-based RF field sensing.

\section{Experimental setup}
\label{sec:2}

\begin{figure}[h]
\includegraphics[width=8cm]{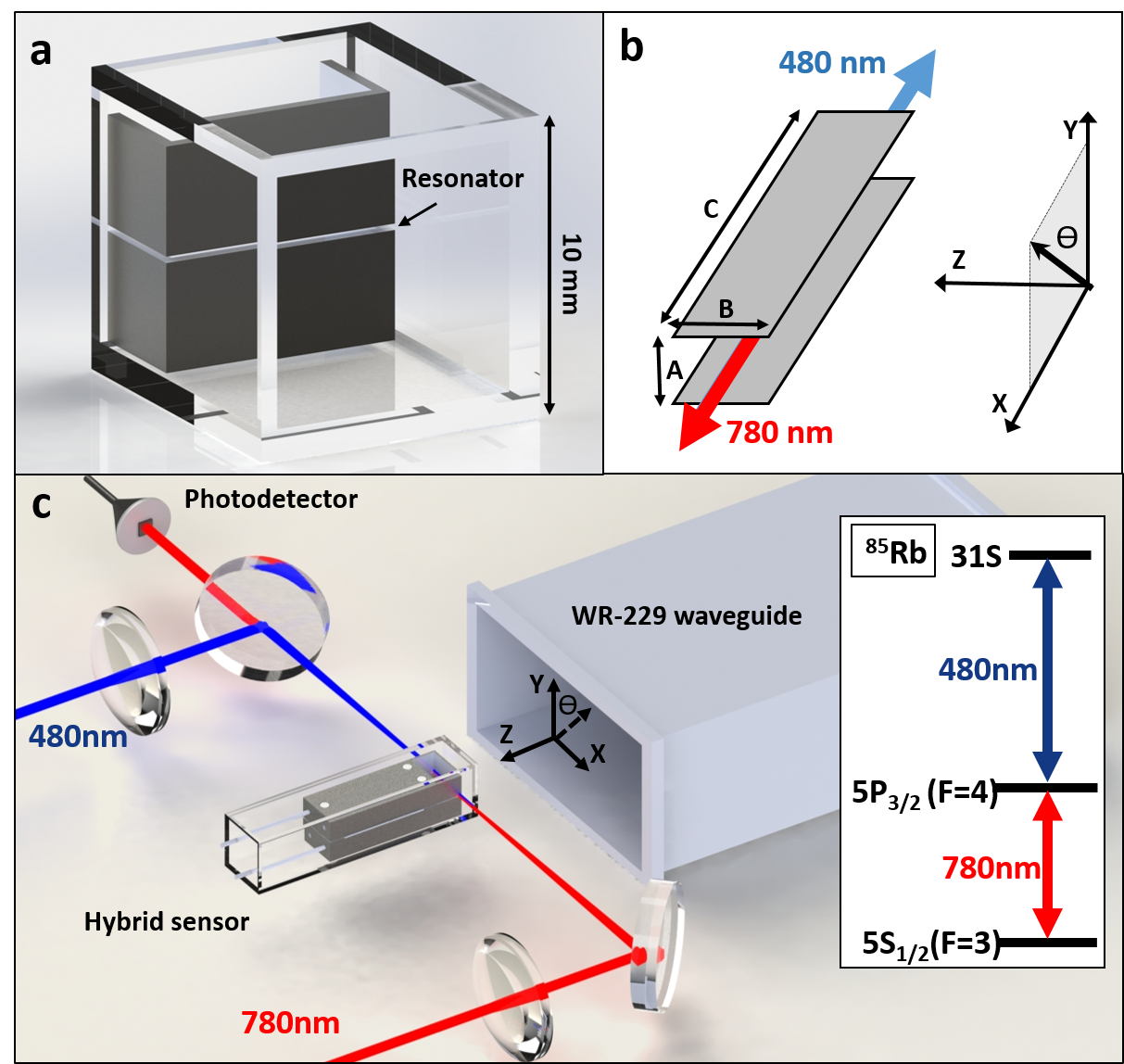}
\caption{(a) Front-end view of a prototype hybrid sensor with resonator embedded in an atomic vapor cell. (b) Illustration of the resonator's electric-field channel with dimensions (A, B, C)=(0.46,0.50,9.0)~mm along $y$, $z$ and $x$, respectively, and geometry of the laser beams used for the optical EIT readout of the RF field. (c) Experimental testing setup for the hybrid sensor, and Rydberg EIT energy-level diagram (inset).}
\label{fig:1}
\end{figure}

The hybrid atomic sensor and its operating principle for RF electric-field sensing is illustrated in Figure~\ref{fig:1}.  Figure~\ref{fig:1}a shows the sensor consisting of an atomic rubidium (Rb) vapor cell with an embedded resonator structure comprised of two stainless steel frames separated by a gap. The impinging electric field resonantly couples into the structure, which compresses and amplifies the RF electric field within the $0.46\times0.50\times9.0$~mm resonator gap volume (geometry shown in Fig.~\ref{fig:1}b). In our measurements, the incident field propagates along the $z$-axis and has a polarization angle $\Theta$ relative to the $y$-axis. The amplified RF electric field induces Rydberg-level shifts or splittings within atoms located inside the field-enhancement gap.  These are optically interrogated using Rydberg EIT, with laser beams directed as illustrated in Fig.~\ref{fig:1}b and transitions as shown in the level diagram in Figure~\ref{fig:1}c. The hybrid sensor in Fig.~\ref{fig:1} has a resonator structure designed for RF detection in the C-band. The cavity geometry is designed to optimally ensure a homogeneous amplification of the field throughout the entire EIT interrogation region, as the beams probe the atomic Rydberg energy levels in the vapor cell. The design affords a 1-cm optical path length through the cell, which is sufficiently long to achieve high absorption and signal-to-noise in the EIT spectra.

We demonstrate field amplification with the hybrid sensor using the experimental setup shown in Fig.~\ref{fig:1}a.  The RF field in the cavity is measured using Rydberg EIT as a high-efficiency non-destructive optical probe for field-induced level shifts of high-lying Rydberg states of $^{85}$Rb atoms within the cavity.  The relevant $^{85}$Rb Rydberg EIT energy-level diagram for optical readout of the field is shown in the inset of Fig.~\ref{fig:1}c.   Two laser beams with wavelengths 780~nm (probe) and 480~nm (coupler) are counter-propagated and overlapped along the resonant channel center, as illustrated in Figs.~\ref{fig:1}b and c.  The 780~nm and 480~nm beams have respective powers of 8~$\mu$W and 40~mW, are both focused to a 70~$\mu$m full-width-at-half-maximum (FWHM), and have linear polarizations along $y$. Rydberg EIT readout is performed by monitoring the 780~nm transmission through the vapor with the laser frequency stabilized to the $^{85}$Rb 5S$_{1/2}$(F=3) to 5P$_{3/2}$(F=4) transition, while the 480~nm laser frequency is scanned linearly across a chosen Rydberg level at a repetition rate of a few Hz.  The 480-nm laser scan is calibrated using a Fabry-P\'erot frequency reference cavity.  For improved signal-to-noise in the optical readout we perform modulation spectroscopy by amplitude modulating the 480~nm beam with a 20~kHz square pulse and demodulating the detected 780~nm signal using a lock-in amplifier.  In our measurements, RF fields are generated using a signal generator, whose output is amplified by a 20~dB amplifier and passed into an open-ended WR-229 waveguide, as shown in Fig.~\ref{fig:1}c. The RF frequency is varied from 2.577 to 5.154~GHz.  The narrow field-enhancement channel of the resonator is positioned approximately 1~cm away from the front face of the waveguide.  In Sec.~\ref{sec:3} the RF field is linearly polarized along the $y$-axis. In Sec.~\ref{sec:4} the RF polarization is varied by rotating the waveguide about the $z$-axis by an angle $\Theta$.

\section{Field amplification}

\label{sec:3}

\begin{figure*}[t]
\includegraphics[width=1\linewidth]{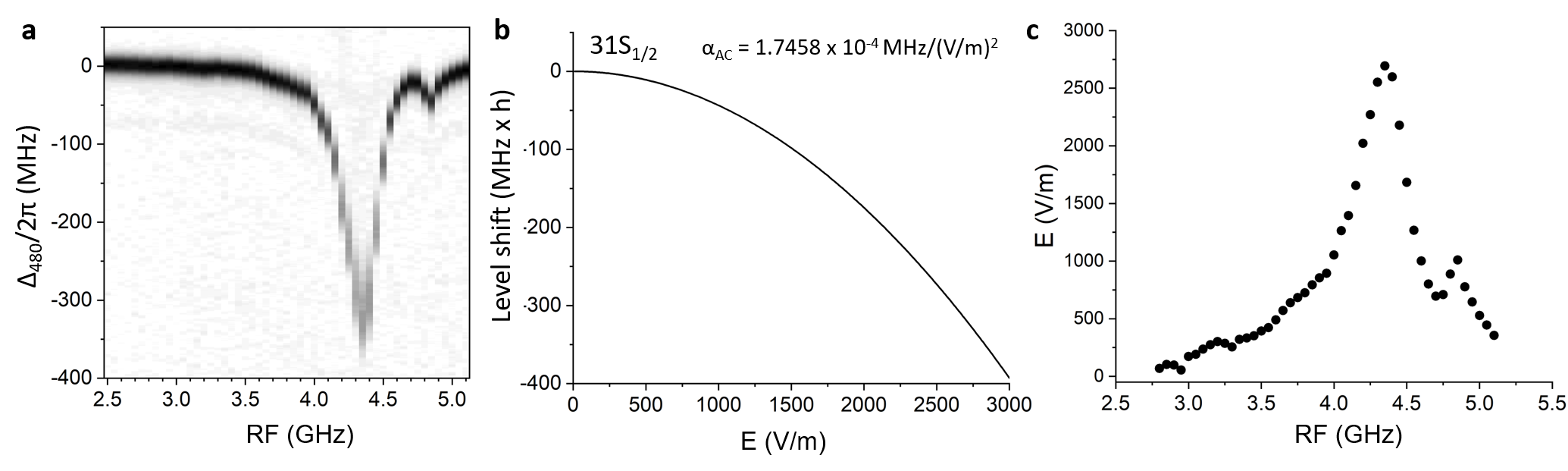}
\caption{Field amplification in a hybrid atomic RF sensor. (a) Measured 31S$_{1/2}$ Rydberg EIT spectra, with 480-nm-laser frequency detuning along the vertical axis, as a function of RF field frequency, at a fixed -10~dBm power injected into the RF amplifier that feeds the microwave transmitter waveguide. (b) Calculated AC Stark shift of the 31S$_{1/2}$ Rydberg state versus RF electric field for an applied 4.35~GHz RF field. (c) Intra-cavity RF electric field $E$ extracted from the measurements in (a) and the mapping function in (b).}
\label{fig:2}
\end{figure*}

To determine RF electric fields we measure AC Stark shifts of the 31S$_{1/2}$ level. The probe and coupler Rabi frequencies at the beam centers are, for our laser powers and beam diameters, calculated to be $\Omega_p = 2 \pi \times 30~$MHz and $\Omega_c = 2 \pi \times 15~$MHz, respectively.
Figure~\ref{fig:2}a shows measured 31S$_{1/2}$ Rydberg-EIT spectra as a function of RF field frequency for a fixed RF power of -10~dBm injected into the amplifier and RF field polarization along $y$.  For the utilized 31S$_{1/2}$ Rydberg level the RF field is far-off-resonant from any dipole-allowed Rydberg transition over the entire tested RF frequency range, and the induced level shift is proportional to the incident RF intensity over a wide RF field range. In Fig.~\ref{fig:2}a, the 480~nm laser frequency is measured relative to the 31S$_{1/2}$ line position with an applied 2.5~GHz RF field; at this frequency the Rydberg-EIT resonance did not exhibit a measurable intra-cavity RF shift.  For increasing RF field frequency, the 31S$_{1/2}$ level begins to substantially shift, with displacements exceeding the EIT line width at around 3.5~GHz and surpassing 300~MHz at the pronounced cavity-induced resonance at 4.35~GHz.  At 4.85~GHz we observe a second, weaker resonance.  The presence of multiple resonances is not unexpected due to the complexity of the metal structure employed here.

The RF electric fields, $E$, for the data in Fig.~\ref{fig:2}a are obtained from the measured AC Stark shifts using the relation $E=(4 \vert \Delta \vert /\alpha_{AC})^{0.5}$, where $\Delta$ is the measured peak line shift and $\alpha_{AC}$ the calculated AC polarizability for the chosen Rydberg state and RF field frequency.  For the Rb 31S$_{1/2}$ state $\alpha_{AC}=1.7458\times 10^{-4}$~MHz/(V/m)$^2$, which varies by less than 0.2\% over the 2.5 to 5.2~GHz frequency range of interest here.  Figure~\ref{fig:2}b shows the calculated shift of the 31$S_{1/2}$ Rydberg state as a function of RF $E$-field.  Figure~\ref{fig:2}c shows the RF $E$-field measured inside the cavity as a function of RF field frequency, obtained from the measured spectra in Fig.~\ref{fig:2}a and the calculation in Fig.~\ref{fig:2}b.  At the 4.35~GHz resonance, we measure a cavity-enhanced field of $E=2690$~V/m with a relative uncertainty of 50~V/m, set by the $\sim$0.1~MHz fitting uncertainty of the peak position.

To determine the cavity-induced electric-field enhancement factor in the hybrid cell, we compare the cavity-enhanced field at 4.35~GHz in Fig.~\ref{fig:2} with a reference measurement of the RF electric field outside of the cavity.  To obtain the latter, the EIT beams are moved from $z=0$, the center of the cavity, by $\Delta z$= -0.9~mm towards the front of the vapor cell.  To obtain a measurable line shift (and field reading) at the reference position outside the cavity, the injected microwave power is increased from -10 to -5~dBm, corresponding to a factor of 3.16 higher incident power and 1.78 higher incident field amplitude, compared to the conditions used in Fig.~\ref{fig:2}. At the reference position we measure a microwave-field-induced AC-Stark shift of $-3.80$~MHz, or a field amplitude of $E=295$~V/m.  Hence, the field enhancement factor afforded by the cavity is $2690\!\times\!1.78/295 = 16.2$, corresponding to a 24.2~dB gain in RF field intensity. In this estimation we neglect that the EIT beams are moved closer by a minute amount (0.9~mm) to the waveguide aperture (which has a short side of 29.1~mm).

\begin{figure}[h]
\includegraphics[width=6cm]{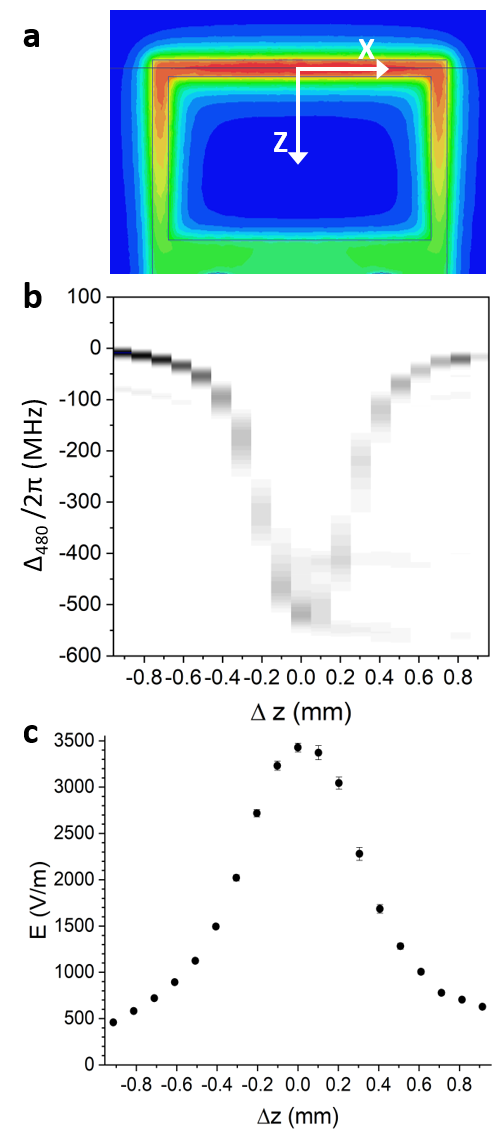}
\caption{(a) Finite-element simulation of the resonant electric-field mode function in the hybrid sensor resonator with a $0.46\times0.50\times9.0$~mm gap for an incident 4.35~GHz microwave field propagating along $z$ with a field of 1~V/m linearly polarized along $y$. The field strength ranges from 1~V/m (blue) to 18.6~V/m (red). (b) 31S$_{1/2}$ EIT spectra versus EIT beam position $\Delta z$ relative to the cavity center for a 4.37-GHz incident microwave field polarized along $y$ at a power of -2~dBm injected into the amplifier. (c) Microwave electric field amplitude versus EIT beam position obtained from the EIT spectroscopic data in (b). Error bars correspond to the fit uncertainties in the peak positions.}
\label{fig:3}
\end{figure}

We benchmark the measured gain in the hybrid sensor against simulations of the electric-field mode function of the
resonance in the cavity for an incident 4.37~GHz field of amplitude 1~V/m linearly polarized along $y$.  Figure~\ref{fig:3}a shows the simulated RF electric field in the gap of the resonator, which is seen to be resonantly enhanced by a factor of about 18.6, corresponding to a 25.4~dB intensity amplification.  This is in excellent agreement with our measured value of 24.2~dB.  The fact that the measured gain is slightly lower may be due to the 0.9-mm change in EIT beam position, or differential refraction and reflection effects of the dielectric materials involved. A systematic study of the effect of dielectric cells on RF electric-field measurements with vapor cells is a topic of interest~\cite{Fan.2015, Anderson.2018} that is beyond the scope of the present work.

In Fig.~\ref{fig:2}a the EIT line width increases from 21.7~MHz at 2.50~GHz to 84.0~MHz at the 4.35~GHz resonance. We attribute this to the mode function inhomogeneity of the 4.35-GHz resonance within the measurement volume. The mode function, a calculated cut of which is shown in Fig.~\ref{fig:3}a,  describes the field distribution that is specific to the resonant modes, and the measurement volume is given by the geometry of the laser beams that constitute the optical EIT field probe. The inhomogeneity of the resonant field mode is strongest at the cavity edges, which are sampled by the wings of the optical beams.  To characterize the field homogeneity in the enhancement region of the hybrid device, we perform a spatial scan of the EIT beams along $z$, relative to the center of the resonator channel at $z=0$.  Figures~\ref{fig:3}b and c show the 31S$_{1/2}$ Rydberg-EIT lines and the averages of the 4.35-GHz microwave field derived from the line shifts, respectively, measured as a function of the spatial shift $\Delta z$ for -2~dBm input power.
The broadening of the EIT line is mostly given by the variation of the mode field along the $z-$direction. The EIT line width increases away from the center of the cavity channel at $z=0$, reaching a maximum at $\Delta z\approx\pm 0.3$~mm at the edges of the 0.5~mm-wide channel.  The asymmetric decrease in the broadening to either side of the channel is in good qualitative agreement with the calculated field map shown in Fig.~\ref{fig:3}a. In Figure~\ref{fig:3}b it is seen that the FWHM of the average-field profile versus $z$ is 0.6~mm, in good agreement with the calculated field profile of the mode.

\section{Polarization sensitivity}
\label{sec:4}
Another feature of hybrid devices is their ability to discriminate between different RF polarizations.  For the resonant structure employed here, the cavity acts as a RF-polarization filter in which only RF fields with a linear polarization component along the $y$-axis of the cavity are coupled into the resonator and field-enhanced within the active measurement volume. The orthogonal RF polarizations are rejected.

\begin{figure}[h]
\includegraphics[width=7cm]{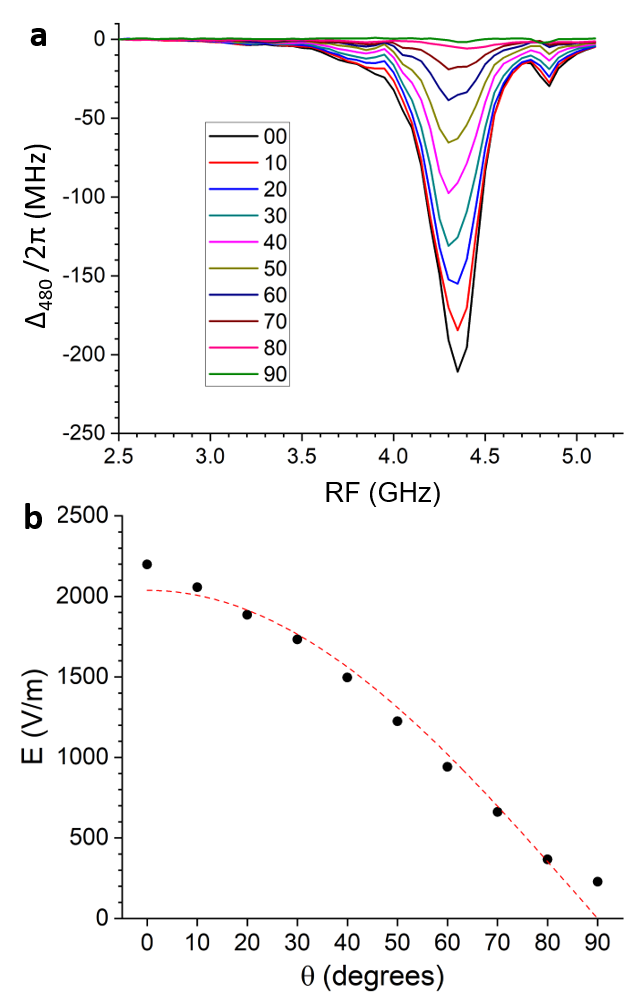}
\caption{(a) 31S$_{1/2}$ line shift versus microwave frequency at a fixed -12~dBm power injected into the amplifier for linearly-polarized microwaves with polarization vector at angle $\Theta$ counter-clockwise from $(x,y) = (0,1)$ in the $xy$-plane (geometry shown in Fig.~\ref{fig:1}). (b) Microwave electric field corresponding to the peak shifts at 4.35~GHz measured in (a) versus $\Theta$.  A cosine fit to the data is given by the dashed line.}
\label{fig:4}
\end{figure}

Figure~\ref{fig:4}a shows experimental 31S$_{1/2}$ Rydberg spectra as a function of RF field frequency at a fixed injected power for different angular alignments $\Theta$ of the RF polarization vector relative to the $y$ axis (see Fig.~\ref{fig:1}a).  This is accomplished by rotating the waveguide in the $xy$-plane counter-clockwise about the $z$-axis from $\Theta = 0^\circ$ (short waveguide axis along $y$) to $90^\circ$ (short axis of waveguide along $x$) in $10^\circ$ increments.  At $\Theta= 0^\circ$ the RF polarization is aligned with the polarization of the resonant cavity-mode function, resulting in maximal coupling into the resonator (black curve in Fig.~\ref{fig:4}a).  As $\Theta$ is increased, the field-induced shift of the EIT peak decreases because the projection of the incident RF electric-field vector onto the $y$-axis is reduced. In Fig.~\ref{fig:4}b we plot of the RF electric field in the cavity, obtained from the measured AC-Stark shifts of the EIT lines shown in Fig.~\ref{fig:4}a, as a function of $\Theta$.
As $\Theta$ is increased, the $y$-component of the field (the component that couples into the resonant field mode of the cavity)
decreases as $\cos(\Theta)$.  A cosine-fit to the data, given by the dashed curve in Fig.~\ref{fig:4}b, confirms this expectation.  It follows that the intensity of a linearly polarized, resonant electromagnetic wave coupled into the resonator and detected by the atoms within the resonator gap has a $\cos^2(\Theta)$-dependence. If the field is measured via an AC Stark shift, as done in the present work, the EIT line shift is $\propto \cos^2(\Theta)$. It is therefore seen that the cavity resonator emulates the functionality of an integrated microwave polarizer, in a way that mimics Malus's law.

In Fig.~\ref{fig:4} we see that the 4.35~GHz field inside the cavity does not reach exactly zero at $\Theta=90^\circ$.  The residual intra-cavity field at $\Theta=90^\circ$ may arise from cavity imperfections, such as slight electrode misalignment and surface-quality issues, as well as from slight polarization imperfections of the field emitted by the waveguide. Such effects are not unexpected for components that are traditionally fabricated at machine tolerances.

\section{Discussion and Conclusion}

A hybrid atomic sensor incorporating a resonant structure for passive radio-frequency field amplification with an atomic vapor cell has been demonstrated. Hybrid detectors with resonator structures either internal or external to the atomic vapor cell and optical readout medium are anticipated to provide improved performance capabilities for Rydberg-atom-based RF field sensors, including high field sensitivity and polarization selectivity for directional field detection.

In the device presented here, the enhancement channel and optical beam geometries were designed to ensure the amplified microwave field remained relatively homogeneous throughout the atomic detection volume, for optical beam sizes that are large enough in diameter so as to avoid interaction-time broadening of the EIT line (given by the average transit time of the thermal atoms through the beams).  For measurement applications requiring EIT line widths narrower than realized with the present geometry, the probe and coupler Rabi frequencies can be reduced to lower the homogeneous EIT line width, and the inhomogeneous field broadening can be reduced by using smaller-diameter laser beams or larger resonant structures with more homogeneous field mode functions, or by a combination of such measures. Minimization of the line broadening will be beneficial, for instance, in hybrid sensors for weak-field microwave measurement based on small transmission changes of very narrow EIT line shapes~\cite{Sedlacek.2012,Kumar.2017}.

We further note that the observation of multiple resonances in Fig.~\ref{fig:2} as well as the general characteristics of eigenmode problems reinforce that it will be possible to engineer hybrid atomic sensors for passive RF field amplification and sensitive detection at selected, application-specific RF frequencies.

\section{Acknowledgements}
This material was supported by Rydberg Technologies, and by the Defense Advanced Research Projects Agency (DARPA) and the Army Contracting Command - Aberdeen Proving Ground (ACC-APG) under Contract number W911NF-15-P-0032.

\end{document}